\documentclass[prb,twocolumn,showpacs]{revtex4}
\usepackage{epsfig,amsmath}

\begin{document}

\title{Theory of antiferromagnetism in the electron-doped cuprate 
superconductors}

\author{Xin-Zhong Yan$^{1,2}$, Qingshan Yuan$^1$, and C. S. Ting$^1$}
\affiliation{$^{1}$Texas Center for Superconductivity, University of 
Houston, Houston, TX 77204\\
$^{2}$Institute of Physics, Chinese Academy of Sciences, P.O. Box 603, 
Beijing 100080, China}

\date{\today}
 
\begin{abstract}
On the basis of the Hubbard model, we present the formulation of antiferromagnetism in electron-doped cuprates using the fluctuation-exchange approach. Taking into account the spin fluctuations in combination with the impurity scattering effect due to the randomly distributed dopant-atoms, we investigate the magnetic properties of the system. It is shown that the antiferromagnetic transition temperature, the onset temperature of the pseudogap formation, the single particle spectral density, and the staggered magnetization obtained by the present approach are in very good agreement with the experimental results. The distribution function in momentum space at very low temperature is observed to differ significantly from that of the Fermi liquid. Also, we find zero-energy peak in the density of states (DOS) of the antiferromagnetic phase. This DOS peak is sharp in the low doping regime, and disappears near the optimal doping where the AF order becomes weak. 
\end{abstract}


\maketitle
\section {Introduction}

The electron-doped cuprate high-temperature superconductors (EDCHTS) such as Nd$_{2-x}$Ce$_x$CuO$_4$ (NCCO) and Pr$_{2-x}$Ce$_x$CuO$_4$ (PCCO) are in the antiferromagnetic (AF) state within a wide doping range at low temperature. \cite{Luke,Kubo,Mang,Thurston} The angle-resolved photoemission spectroscopy (ARPES) measurements\cite{Armitage,Damascelli,Matsui} have revealed the Fermi surface (FS) evolution with doping, providing some microscopic information about the antiferromagnetism in these compounds. On theoretical aspect, much work has been carried out to understand the electronic properties of the EDCHTS by using either Hubbard\cite{Kusko,Kusunose,Senechal,Markiewicz,Kyung,YYT,Yan}or $t-J$ type models.\cite{Yuan} But there are difficulties with the existing theories in describing the AF phase transition and the FS evolution with doping in a consistent manner. To explain the FS evolution and the staggered magnetization as a function of doping, some authors have to assume a doping-dependent Hubbard $U$.\cite{Kusko,Markiewicz} Though the mean-field theory (MFT) yields qualitative explanation for the FS evolution, the calculated AF transition temperature $T_N$ is well known to be too high.\cite{Markiewicz,Yan} In order to reduce the magnitude of $T_N$, the self-consistent random-phase approximation which takes into account the effect of spin fluctuations has been employed, but the obtained $T_N$ is a non-monotonic function with increasing doping concentration $x$,\cite{Markiewicz} contrary to the experimental observation. Therefore, a quantitative understanding of the AF phase transition in EDCHTS remains a challenging issue.

The high-temperature superconductors (HTS) based on cuprates are considered as typical quasi-two-dimensional (Q2D) electron systems. At the early stage after the discovery of the HTS, the antiferromagnetism in such systems has been studied by the random-phase approximation (RPA) with the Hubbard model.\cite{Weng,Shrieffer,Singh,XZY} In this approach the spin fluctuations are treated as perturbation over the mean-field AF state. For the Q2D systems, however, this treatment for the AF state at finite temperature may not be adequate because the number density of the long-wavelength fluctuating spins  diverges logarithmically when the dimensionality approaches the limit 2. It is therefore desirable to treat the AF order and the spin fluctuations on an equal-footing manner. 

Very recently, two of us have extended the fluctuation-exchange (FLEX) approach based on the Hubbard model, which was designed for studying the spin fluctuations in a two-dimensional strongly-correlated electron system, \cite{Bickers} to investigate the antiferromagnetism below $T_N$ in NCCO or PCCO.\cite{YT} In the formalism, the single-particle Green's function, the AF order parameter, and the charge and spin response functions are all treated self-consistently. We found that, in order to quantitatively account for the AF properties, not only the spin fluctuations, but also the impurity scattering effect due to the randomly distributed dopant-atoms are necessary to be taken into consideration. As we know, while the Ce atoms are doped to modulate the carrier density, they should simultaneously introduce disorder which influences the property of electrons in CuO layers. By incorporating the impurity scattering into the FLEX formalism, we have shown, with the Hubbard interaction $U=8$, that the numerically obtained $T_N$, the onset temperature of pseudogap due to spin fluctuations, the single particle spectral density near FS, and the staggered magnetization $m$ in the AF phase as a function of $x$ are all in good agreement with the relevant experimental measurements. 

In this paper, we want to present the detailed FLEX formulation of the AF theory in the EDCHTS which is omitted in Ref.~\onlinecite{YT}, and show some improved and new results as well. By improving the treatment of interlayer coupling and slightly adjusting the parameters, better agreements with experiments can be achieved. As a new result, the single-particle distribution function has been obtained. Its behavior is found to be significantly different from that of the Fermi liquid at low temperature. In addition, the zero-energy peak is found in the density of states (DOS) of the AF phase. This peak is sharp in very low doping regime. The peak becomes suppressed as the doping concentration gets larger and completely disappears near the optimal doping. 

\section{Formalism}

We start with the Hubbard model including the impurity field:
\begin{equation}
H= \sum_{\vec k,\sigma}\xi_{\vec k} c_{k\sigma}^{\dagger}c_{k\sigma} 
+ \frac{U}{N} \sum_{\vec q}n_{\vec q\uparrow} n_{-\vec q\downarrow}
+ \sum_{I}v_{i}n_{I}
\label{Hubbard}
\end{equation}
where $\xi_{\vec k} = \epsilon_{\vec k}-\mu$ with $\epsilon_{\vec k}$ as the tight-binding energy spectrum, $\mu$ the chemical potential, $c_{\vec 
k\sigma}^\dagger$ ($c_{\vec k\sigma}$) is the electron creation (annihilation) operator with spin-$\sigma (= \pm 1$ for up and down spins, respectively) and momentum $\vec k$, $n_{q\sigma}=\sum_{\vec k}c_{\vec 
k\sigma}^\dagger c_{\vec k+\vec q \sigma}$, and $N$ is the total number of the lattice sites. The quantity $v_i$ (assumed to be a constant) in the last term represents the potential acting on the electron with density $n_{I}$ at the impurity site $I$. In NCCO or PCCO, the concentration $N_i$ of the dopant Ce atoms is the same as the electron-doping concentration $x$. The electron system under consideration is a Q2D one with weak interlayer coupling. 

Our main task here is to give a scheme for determining the electronic Green's function $G$. In the normal state, $G$ is related to the self-energy $\Sigma$ via
\begin{eqnarray}
G(\vec k,i\omega_n)= \frac{1}{i\omega_n-\xi_{\vec k}-\Sigma(\vec 
k,i\omega_n)},
\end{eqnarray}
where $\omega_n$ is the fermionic Matsubara frequency. For brevity, we hereafter will use $k \equiv (\vec k, i\omega_n)$ for the argument unless other illustration for special case. The formalism for $G$ is to find out the self-energy $\Sigma$ as a functional of $G$. For constructing the self-energy diagrams, it is the usual way to separate the $U$-interaction term in Eq. (\ref{Hubbard}) into channels of charge density and spin fluctuations,\cite{Esirgen}
\begin{equation}
H_U = \frac{U}{4N} \sum_{\vec q}n_{\vec q} n_{-\vec q}
- \frac{U}{N} \sum_{\vec q}[S^z_{\vec q} S^z_{-\vec q}+S^{+}_{\vec q} 
S^{-}_{\vec q}],
\label{U-int}
\end{equation}
where $n_{\vec q}= n_{\vec q\uparrow}+n_{\vec q\downarrow}$, $S^z_{\vec 
q}= (n_{\vec q\uparrow}-n_{\vec q\downarrow})/2$, $S^+_{\vec q} = 
\sum_{\vec k}c^{\dagger}_{\vec k+\vec q\uparrow}c_{\vec k\downarrow}$ 
($S^-_{\vec q}= S^{+\dagger}_{\vec q}$) are the charge density, longitudinal-spin and transverse-spin fluctuation operators, respectively. Then, diagramming of the self-energy under the FLEX approximation is to piece together these interaction vertices in the right-hand-side (r.h.s.) of Eq. (\ref{U-int}) in each channel. 

\begin{figure} 
\centerline{\epsfig{file=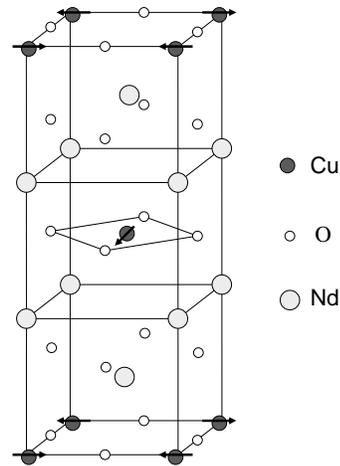,width=6.0 cm}}
%\vskip 2mm
\caption{Crystal structure of Nd$_{2}$CuO$_4$. The arrow shows the spin orientation on the Cu site.}\label{fig1}
\end{figure}

Having obtained Eq. (\ref{U-int}), we also need to take into account the effect due to the interlayer spin coupling. For NCCO or PCCO in which the CuO layers are stacked in a staggered way as shown in Fig. 1, the AF spin coupling between the nearest CuO layers is highly frustrated, and is zero in case of ideal stacking. The local lattice distortions or any other asymmetry may result in a weak AF spin coupling between the nearest layers. In the case of ideal stacking, the coupling between the next-nearest CuO layers is then dominant. We here suppose the ideal stacking and introduce the next-nearest-layer spin coupling Hamiltonian,
\begin{equation}
H_z = \frac{J_z}{2} \sum_{\ell\ell'}[S^z_{\ell} 
S^z_{\ell'}+S^{+}_{\ell} S^{-}_{\ell'}],
\label{zc}
\end{equation}
where $\ell\ell'$-summation is over the next-nearest CuO layers. By adding this $z$-direction coupling to the Hubbard-$U$ term, the interaction is then given by
\begin{equation}
H_{int} = \frac{U}{4N} \sum_{\vec q}n_{\vec q} n_{-\vec q}
-\frac{1}{N} \sum_{\vec q}V(q)[S^z_{\vec q} S^z_{-\vec q}+S^{+}_{\vec 
q} S^{-}_{\vec q}], 
\label{int}
\end{equation}
with
\begin{equation}
V(q) = U-J_z\cos q_z.  \label{V}
\end{equation}
The argument $q_z$ implies that the length unit in $z$-direction is the next-nearest interlayer distance. The weak interlayer coupling means $|J_z| \ll U$. 
Having the constant $J_z$ to mimic the interlayer AF coupling, we then neglect the interlayer electron hopping. The energy $\epsilon_{\vec k}$ in Eq. (\ref{Hubbard}) is hereafter considered as a function of two-dimensional momentum $k$. 

In the following subsections A and B, we construct the self-energy diagrams in the normal and AF states, respectively.

\begin{figure} 
\centerline{\epsfig{file=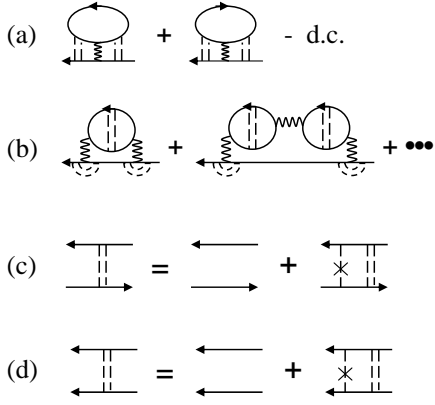,width=6.5 cm}}
%\vskip 2mm
\caption{Diagrams for self-energy due to the charge-density fluctuations in normal state. The wavy line represents the on-site Hubbard $U$. 
The dashed single line with a cross denotes the impurity scattering. (a) 
Lowest order (Hartree) terms, where a double counting (d.c.) without the impurity scattering should be subtracted, (b) ring diagrams representing the charge-density fluctuations, (c) vertex corrections due to the impurity scattering for the particle-hole and (d) particle-particle channels.}\label{fig2}
\end{figure}

\subsection{Normal state}

In the normal state, following the FLEX approximation,\cite{Bickers} we can easily get the self-energy diagrams in the presence of impurities. Firstly, we consider the contribution from charge-density fluctuations represented by the first term in r.h.s. of Eq. (\ref{int}). Its contribution to the self-energy is diagrammatically shown in Fig. \ref{fig2}. The diagrams in Fig. \ref{fig2}(a) come from the Hartree term. The first one includes the diffuson correction due to the particle-hole propagator under the impurity scattering. The second one contains the Cooperon correction. The contribution to the self-energy in the absence of impurity scattering is subtracted because of a double counting and the remaining constant absorbed in the chemical potential. The ring diagrams in Fig. \ref{fig2}(b) represent the charge-density fluctuations with the decoration of the impurity scattering. The bubble in Fig. \ref{fig2}(b) is the charge susceptibility defined by
\begin{equation}
\chi(\vec q,\tau-\tau')\equiv -\frac{1}{2N}\langle T_{\tau}n_{\vec 
q}(\tau)n_{-\vec q}(\tau')\rangle_{bub} 
\label{cs}
\end{equation}
where the subscript $bub$ means the bubble diagram with impurity insertions. The impurity insertions or vertex corrections in the above diagrams are illustrated in Figs. \ref{fig1}(c) and 1(d). They can be expressed by the same formula,
\begin{equation}
\Gamma(q,i\omega_n) = [1-N_iv_i^2X(q,i\omega_n)]^{-1} \label{nvc}
\end{equation}
with
\begin{equation}
X(q,i\omega_n) = \frac{1}{N}\sum_{\vec k}G({\vec k},i\omega_n)G({\vec 
k}+{\vec q},i\omega_n+i\Omega_m),  
\end{equation}
where $q = ({\vec q},i\Omega_m)$ with $\Omega_m$ the bosonic Matsubara frequency. With the vertex correction, in the frequency-space, the charge susceptibility $\chi$ is given by
\begin{equation}
\chi(q) = \frac{1}{\beta}\sum_n\Gamma(q,i\omega_n)X(q,i\omega_n). 
\label{chic} 
\end{equation}
The part of the self-energy due to charge-density fluctuation is then expressed as
\begin{eqnarray}
\Sigma^c 
(k)&=&-\frac{2U}{N\beta}\sum_{q}G(k+q)[\Gamma^2(q,i\omega_n)-1] \cr\cr
& &-\frac{1}{2N\beta}\sum_{q}G(k+q)V^c(q)\Gamma^2(q,i\omega_n) , 
\label{c-sef}
\end{eqnarray}
where
\begin{eqnarray}
V^c(q) = \frac{U^2\chi(q)}{1-U\chi(q)}
\end{eqnarray}
is the polarization potential due to the charge-density fluctuations.

\begin{figure} 
\centerline{\epsfig{file=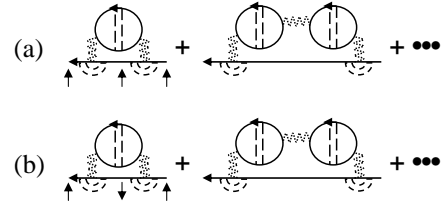,width=6.5 cm}}
%\vskip 2mm
\caption{Self-energy due to the spin fluctuations in the normal state. The point-wavy line represents the interaction $-V(q)$. (a) Ring diagrams representing the longitudinal-spin fluctuations, and (b) transverse-spin fluctuations.}
\label{fig3}
\end{figure}

Analogously, we can treat the spin fluctuations given by the last term in r.h.s. of Eq. (\ref{int}). Figures 3(a) and 3(b) are the self-energy diagrams due to longitudinal-spin and transverse-spin fluctuations, respectively. These diagrams are almost the same as Fig. \ref{fig2}(b) except the bubbles with different meanings. In Fig. \ref{fig3}(a), the bubble is the longitudinal-spin susceptibility defined by
\begin{equation}
\chi^z(\vec q,\tau-\tau')\equiv -\frac{2}{N}\langle T_{\tau}S^z_{\vec 
q}(\tau)S^z_{-\vec q}(\tau')\rangle_{bub}, 
\label{zss}
\end{equation}
while in Fig. \ref{fig3}(b), it is the transverse-spin susceptibility
\begin{equation}
\chi^t(\vec q,\tau-\tau')\equiv -\frac{1}{N}\langle T_{\tau}S^-_{\vec 
q}(\tau)S^+_{\vec q}(\tau')\rangle_{bub}. 
\label{tss1}
\end{equation}
It is easy to see that all three susceptibilities defined above are the same in the normal state. Therefore, we hereafter simply use the notation $\chi(q)$ for them without a superscription. The contribution to the self-energy from the spin fluctuations can then be expressed as
\begin{equation}
\Sigma^s 
(k)=-\frac{1}{N\beta}\sum_{q}G(k+q)[\frac{3}{2}V^s(q)-U^2\chi(q)]\Gamma^2(q,i\omega_n) , \label{s-sef}
\end{equation}
where
\begin{equation}
V^s(q) = \frac{V^2(q)\chi(q)}{1+V(q)\chi(q)} \label{vs}
\end{equation}
is the potential produced by the spin polarization. The term $-U^2\chi(q)$ in Eq. (\ref{s-sef}) represents the subtraction of a double counting in second order of $U$. 

\begin{figure} 
\centerline{\epsfig{file=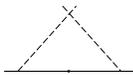,width=2.5 cm}}
%\vskip 2mm
\caption{Self-energy due to the impurity scattering.}\label{fig4}
\end{figure}

The process of the single electron scattering off impurities is shown in Fig. \ref{fig4}. Its contribution to the self-energy is given by
\begin{equation}
\Sigma^i(i\omega_n) = \frac{N_iv_i^2}{N}\sum\limits_{\vec k}G({\vec 
k},i\omega_n).  \label{im}
\end{equation}
The total self-energy is then given by the summation, $\Sigma = 
\Sigma^c+\Sigma^s+\Sigma^i$. Note that by setting $v_i = 0$ and $J_z = 0$, 
$\Sigma$ reduces to the previous result of the FLEX approximation.\cite{Pao,Monthoux}

At temperature above but close to $T_N$, the most important contribution comes from the spin fluctuations represented by Figs. \ref{fig3}(a) and \ref{fig3}(b) since they give rise to the AF instability at $T_N$. The AF instability is characterized by the singularity in $V^s(q)$. At $q = (\vec Q,0)\equiv Q$, the magnitude of $\chi(q)$ reaches its maximum. At $T = T_N$, we have $1+V(Q)\chi(Q)=0$. 

We here emphasize the importance of the interlayer coupling. Without interlayer spin coupling, i.e., $J_z = 0$, the integral in the r.h.s. of Eq. (\ref{s-sef}) will diverge logarithmically at $q = Q$ and finite temperature, which means that no AF transition occurs at finite temperature for two-dimensional (2D) systems. Though $J_z$ is small, it controls the overall magnitude of the AF transition temperature $T_N$. 

The interlayer spin coupling becomes significant only at the singular point of $V^s(q)$. The Green's function $G$ and the susceptibility $\chi$ very weakly depend on the weak interlayer coupling. By neglecting the weak $q_z$-dependence in $G$ and 
$\chi$, we can then integrate out Eq. (\ref{s-sef}) over $q_z$ immediately, and reduce the three-dimensional problem to a 2D one. Since the delicate $q_z$-dependence of $V^s(q)$ comes from the denominator in Eq. (\ref{vs}), we neglect the small $J_z$ in $V^2(q)$ in the numerator. By integrating over $q_z$, $V^s(q)$ is replaced by 
\begin{equation}
\bar V^s(q) = \frac{U^2\chi(q)}{\sqrt{[1+U\chi(q)]^2-J^2_z\chi^2(q)}}. 
\label{vsa}
\end{equation}
This result is just the average of $V^s(q)$ over $-\pi<q_z<\pi$. In our numerical calculation, $|J_z|/U < 2\%$, the above approximation should be reasonable.

\begin{figure} 
\centerline{\epsfig{file=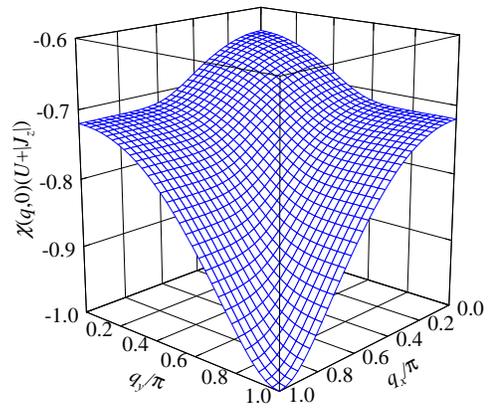, width=6.5 cm}}
%\vskip 2mm
\caption{(Color online) Static spin susceptibility $\chi(\vec 
q,0)\times(U+|J_z|)$ at $T = T_N$ and $x = 0.1$. }\label{fig5}
\end{figure}

Figure \ref{fig5} shows a numerical result for $\chi(\vec q,0)$ at $T = T_N$ for doping concentration at $x = 0.1$, and it indicates that the maximum of $|\chi(\vec q,0)|$ appears at $(Q_x,Q_y) = (\pi,\pi)$. Our numerical calculation also shows that such behavior for $\chi(\vec q,0)$ remains unchanged in the whole doping range investigated here. This is different from the hole-doped case where the maximum appears at some incommensurate wavevectors close to $(\pi,\pi)$. The reason is that the Fermi surface can be nested at $(Q_x,Q_y) = (\pi,\pi)$ for the electron doped sample within a large doping range. For the hole doped sample, the Fermi surface cannot be nested at $(Q_x,Q_y) = (\pi,\pi)$ unless in the very low doping regime. As for the $z$-component, $Q_z = 0$ or $\pi$ depends on the sign of $J_z$. Under our approximation where $V^s(q)$ is replaced by $\bar V^s(q)$, the final result is independent of $Q_z$. In NCCO and PCCO, $Q_z = 0$ is observed, which means a negative $J_z$ or ferromagnetic interlayer coupling.

\subsection{AF phase}

Below $T_N$, the system is in the AF state that is characterized by the staggered magnetization $m$. Now, the operator $S^z_{\vec Q} \to -Nm$ is a macroscopic quantity. The term of $\vec q = \vec Q$ in the spin interaction in Eq. (\ref{int}) should be singled out, and be treated as (up to a constant)
\begin{equation}
-\frac{V(Q)}{N}S^z_{\vec Q}S^z_{-\vec Q}\approx \Delta\sum_{\vec 
k}(c_{\vec k\uparrow}^\dagger c_{\vec k+\vec Q \uparrow}-c_{\vec 
k\downarrow}^\dagger c_{\vec k+\vec Q \downarrow})  
\label{MFT}
\end{equation}
where $\Delta = V(Q)m/2$ is the order parameter. Though Eq. (\ref{MFT}) looks like the mean-field result, we here also take into account simultaneously the fluctuation terms $\vec q \ne \vec Q$ in the interaction. Because of Eq. (\ref{MFT}), we need to redefine the Green's function as a $2\times 2$ matrix 
\begin{equation}
\hat G(\vec k,\tau-\tau')\equiv -\langle T_{\tau}\psi_{\vec 
k\sigma}(\tau)\psi^{\dagger}_{\vec k\sigma}(\tau')\rangle 
\label{GF}
\end{equation}
with $\psi^{\dagger}_{\vec k\sigma}\equiv (c^{\dagger}_{\vec 
k\sigma},\sigma c^{\dagger}_{\vec k+\vec Q \sigma})$. Since the potential 
($\pm\Delta$) in Eq. (\ref{MFT}) for opposite spin electrons has opposite sign, $\hat G$ so defined is independent of the spin. In the Matsubara frequency space, 
in terms of the self-energy $\hat\Sigma(k)$ which is accordingly a $2\times 2$ matrix, $\hat G(k)$ is written as
\begin{equation}
\hat G(k) = [i\omega_n-\xi_0(\vec k)-\xi_3(\vec 
k)\sigma_3-\hat\Sigma(k)]^{-1},
\label{GFs}
\end{equation}
where $\xi_{0(3)}(\vec k)=(\xi_{\vec k}\pm \xi_{\vec k+\vec Q})/2$, $\sigma_3$ is the Pauli matrix, and the scalar quantity $i\omega_n-\xi_0(\vec k)$ is actually multiplied by the unit matrix $\sigma_0$ that is dropped out for brevity. The four elements of $\hat G$ have the following relationship:
\begin{eqnarray}
G_{22}(k)&=&G_{11}(k+Q) \cr\cr
G_{21}(k)&=&G_{12}(k)=G_{12}(k+Q). 
\end{eqnarray}
Similarly, the elements of $\hat \Sigma$ have the following relationship:
\begin{eqnarray}
\Sigma_{22}(k)&=&\Sigma_{11}( k+Q) \cr\cr
\Sigma_{21}(k)&=&\Sigma_{12}(k)=\Sigma_{12}(k+Q). 
\end{eqnarray}
By definition, the order parameter $\Delta$ can be written as
\begin{equation}
\Delta=-\frac{V(Q)}{N\beta}\sum_{k}G_{12}(k) \label{op}
\end{equation}
where $\beta$ is the inverse of the temperature $T$. From Eq. (\ref{MFT}) with the impurity influence, the off-diagonal self-energy $\Sigma_{12}(k)$ is obtained as in Fig. 6.
 
\begin{figure}
\centerline{\epsfig{file=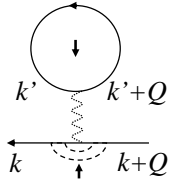, width=2.5 cm}}
\caption{Off-diagonal self-energy $\Sigma_{12}$.}\label{fig6}
\end{figure}

For the diagonal self-energy, we need to extend the results shown in Figs. \ref{fig2}-\ref{fig4} for the normal state to that of the AF state. Since the diagonal and off-diagonal parts of the self-energy have a kind of harmonious relationship, they cannot be treated independently. Here, two points need to be noted: (1) Except the diagram shown in Fig. \ref{fig6}, are there any others that should be included in $\Sigma_{12}(k)$ under the ring approximation for the diagonal part? The transverse-spin fluctuations corresponding to Fig. \ref{fig3}(b) should include the collective modes, especially the Goldstone mode (GM) that is a consequence of the broken symmetry. The equation determining the GM should be consistent with the gap equation (\ref{op}). This consistency gives rise to a constraint on the form of $\Sigma_{12}$. We will show that Fig. \ref{fig6} is the only graph for $\Sigma_{12}$ within the ring approximation for the diagonal self-energy. (2) What should be the form of the impurity insertion? The insertion of impurity scatterings in the transverse-spin fluctuation is between the opposite spin electrons, while in 
Fig. \ref{fig6} it is between the same spin electrons. In order to ensure the consistency between GM and Eq. (\ref{op}), these insertions should be the same, and the scattering processes including $G_{12}$ will be excluded. In the following, we give detailed explanations to these two points with and without impurity scatterings. 

1. Without impurity scattering (i.e., $v_i = 0$). In the AF state, since the transverse-spin fluctuations of momenta $\vec q$ and $\vec q+\vec Q$ are correlated, we need to consider a matrix $\chi^t$ for the transverse-spin susceptibilities. For the purpose, we denote the transverse-spin operators by
\begin{equation}
S_{\mu}(\vec q)= \sum_{\vec k}c^{\dagger}_{\vec k\downarrow}c_{\vec 
k+\vec Q_{\mu}\uparrow}
\end{equation}
where $\vec Q_{\mu} = 0~(\rm for~\mu =1)$, or $\vec Q (\rm for~\mu =2)$. The matrix elements are then defined by
\begin{equation}
\chi^t_{\mu\nu}(\vec q, \tau-\tau')=-\frac{1}{N}\langle 
T_{\tau}S_{\mu}(\vec q,\tau)S^{\dagger}_{\nu}(\vec q,\tau')\rangle_{bub}.
\end{equation}
After Fourier transform from $\tau$ to $i\Omega_m$, they are given by
\begin{eqnarray}
\chi^t_{11}(q) =\frac{1}{N\beta}\sum_k 
[G_{11}(k)G_{11}(k+q)-G_{12}(k)G_{12}(k+q)] \nonumber\\
\chi^t_{12}(q) =\frac{1}{N\beta}\sum_k 
[G_{11}(k)G_{12}(k+q)-G_{12}(k)G_{11}(k+q)] \nonumber
\end{eqnarray}
and $\chi^t_{22}(q)=\chi^t_{11}(q+Q)$, $\chi^t_{21}(q)=\chi^t_{12}(q)$. 
In terms of $\hat \chi^t(q)$, we can then express the transverse-spin polarization potential $\hat V^t(q)$ [stemming from a similar bubble summation as in Fig. \ref{fig3}b] as
\begin{equation}
\hat V^t(q)= [1+ \hat V(q)\hat\chi^t(q)]^{-1}\hat V(q)\hat\chi^t(q)\hat 
V(q), \label{pafvs}
\end{equation}
which is a $2\times 2$ matrix too. Note that $\hat V(q)$ is the scalar 
$V(q)$ as given by Eq. (\ref{V}) if $Q_z = 0$, or a matrix $\hat V(q) = U - 
\sigma_3J_z\cos q_z$ if $Q_z = \pi$. Here, an important point is the divergence of $\hat V^t(q)$ at $q = Q$ [where $\chi^t_{12}(Q)=0$], that is
\begin{equation}
1+V_{11}(Q)\chi^t_{11}(Q) = 0 \label{GM} 
\end{equation}
which means the existence of the Goldstone mode. To see the constraint on the off-diagonal diagrams, we express Eq. (\ref{GM}) in the following form. Since $V_{11}(q) = V(q)$, we drop the subscripts in the following. Using the relationship between $G$ and $\Sigma$, we have
\begin{eqnarray}
V(Q)\chi^t_{11}(Q) &=& 
\frac{V(Q)}{N\beta}\sum_k[G_{11}(k)G_{22}(k)-G^2_{12}(k)] \nonumber\\
&=& \frac{V(Q)}{N\beta}\sum_k G_{12}(k)/\Sigma_{12}(k) \nonumber\\
&=& -1.
\end{eqnarray}
By comparing the last equation above with the gap equation (\ref{op}), one gets 
\begin{equation}
\Sigma_{12}(k) = \Delta.
\end{equation} 
In other words, if one deals with $\Sigma_{12}(k)$ improperly, Eq. (\ref{GM}) cannot be satisfied. Therefore, under the ring approximation for the diagonal self-energy, $\Delta$ is the only term for $\Sigma_{12}(k)$. 

\begin{figure}
\centerline{\epsfig{file=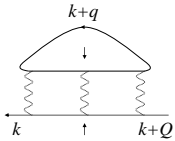, width=2.5 cm}}
\caption{Contribution to the off-diagonal self-energy $\Sigma_{12}$ in RPA. Under the present approximation, this term should be excluded.}\label{fig7}
\end{figure}

It is suggestive to compare the above result with that from RPA. Within RPA the contribution from the coupling between the transverse spin fluctuations 
$S_1(\vec q)$ and $S_2^{\dagger}(\vec q)$ is taken into account in 
$\Sigma_{12}(k)$. Such a diagram is shown in Fig. \ref{fig7}. However, this contribution should be excluded under the present approximation. Accordingly, the similar contribution from such coupling to the diagonal part should be excluded as well. The diagonal self-energy due to the transverse-spin fluctuations is given by 
\begin{equation}
\Sigma^{t}(k)=-\frac{1}{N}\sum_qG(k+q)[V^{t}_{11}(q)-U^2\chi^t_{11}(q)],
\label{tslf}
\end{equation}
where again the term $-U^2\chi^t_{11}(q)$ represents the subtraction of the double counting in the second order of $U$.

2. In the presence of impurity scatterings. The off-diagonal self-energy $\Sigma_{12}$ represented by the diagram in Fig. \ref{fig6} is written as, 
\begin{equation}
\Sigma_{12}(k)=\Delta\Gamma(Q,i\omega_n). \label{off1}
\end{equation}
with $\Gamma(Q,i\omega_n)$ as vertex correction due to impurity scatterings. Note that the spins on the two sidelines of the vertex $\Gamma(Q,i\omega_n)$ in Fig. \ref{fig6} are the same. In the AF state, besides the Green's function $G_{11}$, one may consider the insertions including $G_{12}$ in $\Gamma(Q,i\omega_n)$. If $G_{12}$ exists in the vertex correction in Fig. \ref{fig6}, it must appear in both sides of an impurity line as a couple. Since their spins are the same, one obtains a term of $G_{12}(k)G_{12}(k)$ in the $k$-summation in $X(Q,i\omega_n)$. In the following, however, we show that these terms including $G_{12}$ should be excluded in order to maintain the consistence between the Eq. (\ref{off1}) and the Goldstone mode. Analogous to what is done above, we consider the transverse-spin polarization potential $\hat V^t(q)$ in the presence of the impurities. The potential $\hat 
V^t(q)$ is given by Eq. (\ref{pafvs}) but with the vertex correction to the susceptibility included. We here only consider the case of $q = Q$, where $\hat V^t(q)$ is singular. It means  
\begin{eqnarray}
V(Q)\chi^t_{11}(Q)
&=& 
\frac{V(Q)}{N\beta}\sum_k\Gamma_{11}(Q,i\omega_n)[G_{11}(k)G_{22}(k)-G^2_{12}(k)]\nonumber\\
&=& \frac{U}{N\beta}\sum_k 
G_{12}(k)\Gamma_{11}(Q,i\omega_n)/\Sigma_{12}(k) \nonumber\\
&=& -1,\nonumber
\end{eqnarray}
where $\Gamma_{11}(Q,i\omega_n)$ is the impurity insertion in the susceptibility $\chi^t_{11}(Q)$. Comparing this result with Eq. (\ref{op}), we have $\Sigma_{12}(k)=\Delta\Gamma_{11}(Q,i\omega_n)$. On the other hand, from Eq. (\ref{off1}), we get $\Gamma_{11}(Q,i\omega_n)=\Gamma(Q,i\omega_n)$. However, 
$\Gamma_{11}(Q,i\omega_n)$ is the impurity insertion with the opposite spins in the two legs of the ladder. If there is $G_{12}$, one should get a term $-G_{12}(k)G_{12}(k)$ in $\Gamma_{11}(Q,i\omega_n)$, whose sign is different from that in $\Gamma(Q,i\omega_n)$. Therefore, those terms including $G_{12}$ should be excluded from the vertex corrections. 

\begin{figure}
\centerline{\epsfig{file=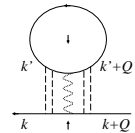, width=2. cm}}
\caption{A form for the impurity insertions in the off-diagonal self-energy $\Sigma_{12}$. Under the present approximation, this vertex correction is not consistent with the Goldstone mode, and should be excluded.}\label{fig8}
\end{figure}

The approximation for the off-diagonal self-energy is very constrained by the consistence between the gap equation and the Goldstone mode. For example, shown in Fig. \ref{fig8} is a possible case for the impurity insertions in $\Sigma_{12}$. But this kind of vertex correction is not consistent with the Goldstone mode under the present approximation, and therefore should be excluded.

With these two points clarified, we can then give the expression for the self-energy in the AF state. We start with the expression for the vertex corrections. As discussed above, the functional form of the vertex correction is given by Eq. (\ref{pafvs}) but with the Green's function $G$ understood to be $G_{11}$, 
$\Gamma(q,i\omega_n) = [1-N_iv_i^2X(q,i\omega_n)]^{-1}$, 
with
\begin{equation} 
X(q,i\omega_n) = \frac{1}{N}\sum_{\vec k}G_{11}(k)G_{11}(k+q). 
\nonumber
\end{equation}
With the vertex correction, the longitudinal $\chi^z$ and transverse 
$\chi^t$ spin susceptibilities are given by
\begin{eqnarray}
\chi^z(q) = \frac{1}{\beta}\sum_n\Gamma(q,i\omega_n)X_{+}(q,i\omega_n), 
\label{chiz}\\
\chi^t_{11}(q) 
=\frac{1}{\beta}\sum_n\Gamma(q,i\omega_n)X_{-}(q,i\omega_n), \label{ch11}\\
\chi^t_{12}(q) = 
\frac{1}{\beta}\sum_n\Gamma(q,i\omega_n)X_o(q,i\omega_n)\Gamma(q+Q,i\omega_n), \label{ch12}
\end{eqnarray}
with
\begin{eqnarray}
X_{\pm}(q,i\omega_n) = \frac{1}{N}\sum_{\vec k}[G_{11}(k)G_{11}(k+q)\pm 
G_{12}(k)G_{12}(k+q)] \nonumber\\ 
X_{o}(q,i\omega_n)=\frac{1}{N}\sum_{\vec 
k}[G_{11}(k)G_{12}(k+q)-G_{12}(k)G_{11}(k+q)]. \nonumber
\end{eqnarray}
The expression for $\Sigma_{11}$ reads
\begin{eqnarray}
\Sigma_{11}(k)&=&-\frac{2U}{N\beta}\sum_{q}G_{11}(k+q)[\Gamma^2(q,i\omega_n)-1]+ 
\Sigma_i(i\omega_n) \cr\cr
& &-\frac{1}{N\beta}\sum_{q}G_{11}(k+q)V_{\rm 
eff}(q)\Gamma^2(q,i\omega_n) , \label{sef}
\end{eqnarray}
with $V_{\rm eff}(q) = V^c(q)/2+ V^z(q)/2 + V^t_{11}(q)-U^2\chi^t_{11}(q)$ 
and
\begin{eqnarray}
V^c(q) = \frac{U^2\chi^z(q)}{1-U\chi^z(q)}, \label{vc}\\
V^z(q) = \frac{V^2(q)\chi^z(q)}{1+V(q)\chi^z(q)}, 
\label{vz}\\
\Sigma_i(i\omega_n) = \frac{N_iv_i^2}{N}\sum\limits_{\vec 
k}G_{11}({\vec k},i\omega_n).  \label{tm}
\end{eqnarray}
The term $V^c(q)$ is due to the charge-density fluctuations, while the $V^z(q)$-term stems from the longitudinal-spin fluctuations. At low $T$ with finite $\Delta$, since $V^t_{11}(q)$ diverges at $q=Q$, the most important contribution to the diagonal self-energy comes from the transverse-spin fluctuations. The corresponding diagram is an extension of Fig. \ref{fig3}(b). The other elements of $\hat\Sigma(k)$ are given by $\Sigma_{22}(k)=\Sigma_{11}(k+Q)$, and $\Sigma_{21}(k)=\Sigma_{12}(k)$.  

Here again, we discuss the interlayer coupling $J_z$ that appears in 
$V^z(q)$ and $V^t_{11}(q)$. For $V^z(q)$, since its sensitive dependence on $J_z$ comes from the denominator that vanishes at $T_N$ in r.h.s. of Eq. (\ref{vz}), we neglect the 
$J_z$-dependence in the numerator and average it over $-\pi<q_z<\pi$ as what have been done for normal state. By the same consideration, we can deal with $V^t_{11}(q)$. But the situation for $V^t_{11}(q)$ is a little more complicated, we consider it in Appendix. In our previous calculation,\cite{YT} the interlayer spin coupling was taken into account artificially by introducing a weak $q^2_z$-dependence in these denominators, which corresponds to the expansion of $\cos q_z \approx 1-q^2_z/2$ in the denominators to order of $q^2_z$ but neglecting the $J_z$-dependent constant term. The physics is that the long-wavelength spin-fluctuations should be treated as three-dimensional, which is the same as present consideration. 

\section{Numerical results}

In the present calculation, we take the nearest-neighbor (NN) hopping
$t$ as the energy unit (i.e., $t=1$) and choose
$t' = -0.25$ (next NN hopping), $t'' = 0.1$ (third NN hopping), $U = 6.6$,
$J_z = -0.1$, and $v_i = 1.55$. The adopted values for $t'$, $t''$ and $U$ are about the same as those used in the literature.\cite{Kusko,Senechal} With the recently developed algorithms for dealing with the summation over the Matsubara frequency and the Fourier transform of the effective interaction with long-wavelength singularity,\cite{Yan2} we have numerically obtained the self-consistent solution to the Green's function. In this section, we present the numerical results for the physical quantities.

\subsection{Normal state} 
\begin{figure}
\centerline{\epsfig{file=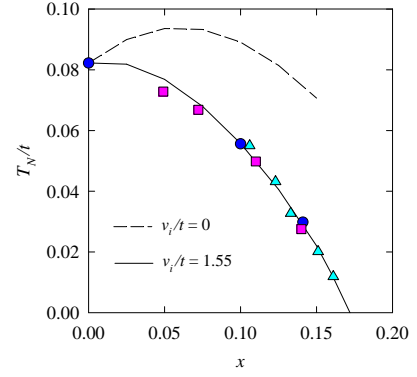, width=5.5 cm}}
\caption{(Color online) $T_N$ as a function of doping concentration $x$. 
Solid ($v_i/t = 1.55$) and dashed ($v_i = 0$) curves are the present calculated results. The solid circles,\cite{Luke} triangles,\cite{Kubo} 
and squares \cite{Mang} are the experimental data.} \label{fig9}
\end{figure}

In Fig. \ref{fig9}, the numerical results (solid and dashed curves) are presented for $T_N$ as functions of $x$. For comparison, the experimental data for NCCO \cite{Luke,Kubo,Mang} are depicted by scaling the maximum $T_N =255$ K to $0.0822t$ (which means $t \approx 270$ meV). The present result for $T_N$ is about one order of magnitude smaller than that from the MFT.\cite{Markiewicz,Yan} Even though the FLEX approach without consideration of the impurity scattering ($v_i = 0$) gives the overall lower $T_N$, its non-monotonic behavior does not reflect the feature of the experimental data. In contrast, the calculation with impurity strength $v_i/t = 1.55$ fits very well the experimental data. In our calculation, the overall magnitude of $T_N$ is controlled by $U$ and $J_z$; larger $U$ and $J_z$ lead to higher $T_N$. With increasing $v_i$ from zero, the curve $T_N(x)$ is suppressed gradually from the non-monotonic result toward a monotonic decreasing (as a function of $x$) one.  

\begin{figure}
\centerline{\epsfig{file=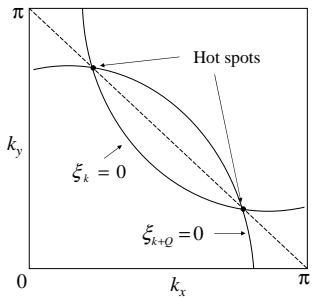, width=5. cm}}
\caption{Sketch of the hot spots in the first quadrant of the Brillouin zone.}\label{fig10}
\end{figure}

In order to understand how the impurities suppress the AF transition, we firstly analyze the criterion $1+U\chi(Q)=0$ in the absence of impurities. In this case $\Gamma = 1$, the spin susceptibility $\chi$ is simply determined by $X(q,i\omega_n)$. We note that at $q = Q$, a large contribution to $\chi(Q)$ stems from the small regions around the hot spots $\vec k_h$ which are the cross points of $\xi_{\vec k} = 0$ and $\xi_{\vec k +\vec Q} = 0$ as shown in Fig. \ref{fig10}. For the free particles, one gets $G^{(0)}({\vec k_h},i\omega_n)G^{(0)}({\vec k_h}+{\vec Q},i\omega_n)= -\omega_n^{-2}$ since the energy $\xi_{\vec k_h}=\xi_{\vec k_h+\vec Q} = 0$. In the present case the electrons interact via spin fluctuations, $V^s(q)$ has a divergent peak at $q = (\vec Q, 0)$ in the limit $T \to T_N$ from above. Considering only the predominant contribution from long-wavelength spin fluctuations, we then have approximately $\Sigma(k) \approx \gamma^2G(\vec k + \vec Q, i\omega_n)$ with $\gamma^2 = 
-3\sum_{q}V^s(q)/2N\beta$ where the ${\vec q}$-summation runs over a small region around ${\vec Q}$. By this approximation, $G$ can be obtained analytically. Especially, at the hot spots, we have $G({\vec k_h},i\omega_n)\approx 2/i\omega_n(1+\sqrt{1+4\gamma^2/\omega_n^2} )$, which implies a reduction in the magnitude of $G$. Even at large doping where hot spots vanish, the magnitude of $G$ can be reduced by the spin fluctuations as well. Thus, the magnitude of $X$ and thereby $|\chi(Q)|$ are reduced. Since $|\chi(Q)|$ increases with decreasing $T$, one therefore obtains a lower $T_N$ than that from the MFT. In the presence of impurities, the magnitude of 
$\chi(Q)$ is further decreased by $\Gamma$ ($|\Gamma| < 1$), which leads to a further lowering of $T_N$. 
 
Here we also wish to give the reason why the FLEX result for $T_N$ at $v_i=0$ is a non-monotonic function of $x$ in the absence of impurity scattering. Again, we consider the criterion $1+U\chi(Q)=0$ that determines $T_N$. As mentioned above, a large contribution to $\chi(Q)$ stems from the small regions around the hot spots $\vec k_h$. Let us look at the two hot spots in the first quadrant of the Brillouin zone. With increasing $x$ from zero, these two hot spots move toward each other. Meanwhile, between the two hot spots, the two curves $\xi_{\vec k} = 0$ and $\xi_{\vec k +\vec Q} = 0$ approach each other too. Therefore, the inter-region enclosed by the curves between the two hot spots begins to give considerable contribution to $\chi(Q)$. This additional contribution enhances the magnitude of $\chi(Q)$, and thereby leads to increasing of $T_N$ with $x$. On the other hand, at large doping limit, there are no hot spots. With increasing doping, the two curves depart from each other, resulting in a decreasing of $T_N$. Therefore, there is a maximum in $T_N$ at certain $x$ within the FLEX calculation. With the impurity scattering, however, the density $N_i$ increases as the same as $x$, reducing the magnitude of the vertex function $\Gamma$. Consequently, at small $x$, $T_N$ cannot increase as much as that predicted by FLEX. In the present case, with $v_i = 1.55$, $T_N$ monotonically decreases with $x$.  

\begin{figure}
\centerline{\epsfig{file=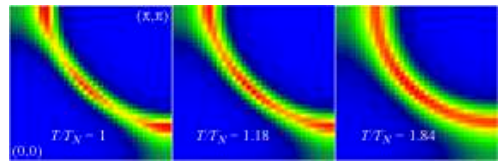, width=7. cm}}
\caption{(Color online) Occupied spectral density around the chemical potential at temperatures $T/T_N = 1$, 1.18, and 1.84, at doping concentration $x = 0.05$.}\label{fig11}
\end{figure}

It is expected that spin fluctuation effect may be observable above but close to $T_N$. For example, the effect can be clearly reflected in the spectral density which is defined as
\begin{equation}
A(\vec k,E)= -\frac{1}{\pi}{\rm Im}G(\vec k,E+i0^+). 
\end{equation}
The variation of the spectral density with temperature may be detected by the ARPES experiment. By use of the Pad\'e approximation,\cite{Vidberg} the analytical continuation for the Green's function and thereby the function $A(\vec k,E)$ can be obtained. Fig. \ref{fig11} presents the occupied spectral density $A^<(\vec k,E)) = f(E)A(\vec k,E)$ (with $f$ the Fermi distribution function) around the chemical potential integrated within the energy window of $-0.02 t < E < 0.02 t$ at $T/T_N =1$, 1.18 and 1.84, and at $x = 0.05$. At this doping, far below $T_N$, the spectral intensity is expected to mainly concentrate around the points $(\pi,0)$ and $(0,\pi)$. However, although there is no AF order, the Fermi surface splits into three parts at $T$ above but close to $T_N$. The intensity loses at the hot spots, which is due to the formation of the pseudogap caused by spin fluctuations. When $T$ above $T_N$ continuously increases, the spin fluctuations become weak and the pseudogap diminishes at the hot spots. As a result, the Fermi surface becomes a continuous curve as seen in the case $T/T_N=1.84$.

\begin{figure}
\centerline{\epsfig{file=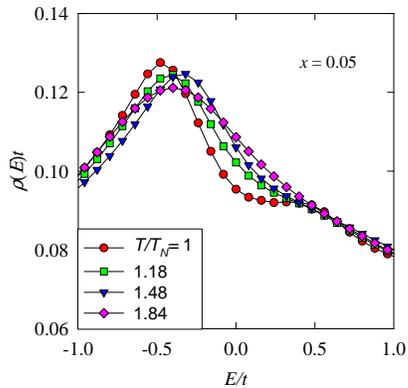, width=5.5 cm}}
\caption{(Color online) Density of states at $x = 0.05$ and $T/T_N = 1$, 1.18, 1.48, 1.84.}\label{fig12}
\end{figure}

The pseudogap can be also reflected in the density of states (DOS) which is given by $\rho(E)=(1/N)\sum_{\vec k}A(\vec k,E)$. Shown in Fig. \ref{fig12} are the results for DOS at $x = 0.05$ and various temperatures. As the temperature decreases from high value to $T_N$, DOS changes gradually. Especially, around $E = 0$, a visible depression of DOS due to the pseudogap formation is seen as $T \to T_N$. The onset temperature $T^*$ for the depression at this doping is about $1.7T_N$. 

\begin{figure}
\centerline{\epsfig{file=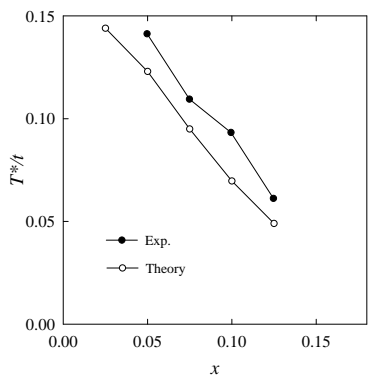, width=5. cm}}
\caption{Pseudogap onset temperatures $T^*$ as a function of doping concentration $x$. The present result (theory) is compared with the experimental data of Ref.~{\onlinecite{Onose}}.}\label{fig13}
\end{figure}

In Fig. \ref{fig13}, $T^*$ is shown as a function of doping concentration, and is compared with the experimental data extracted from measurements on optical conductivity.\cite{Onose} Clearly, our calculated $T^{*}$ is consistent with the experimental data. Similar result on $T^{*}$ has also been obtained by Kyung {\it et al.} \cite{Kyung} using a different approach.

So far we showed the effects of spin fluctuations and impurity scattering on the properties of the normal state. In the next subsection, we present the results in the AF state. 

\subsection{AF phase}

\begin{figure} 
\centerline{\epsfig{file=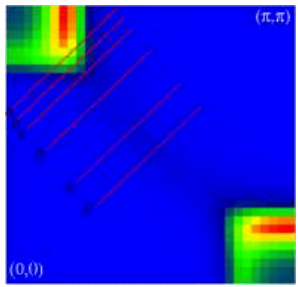, width=4.5 cm}}
\caption{(Color online) Occupied spectral density around the chemical potential at $T/t = 0.003$ and $x = 0.1$.}\label{fig14}
\end{figure}

\begin{figure}
\centerline{\epsfig{file=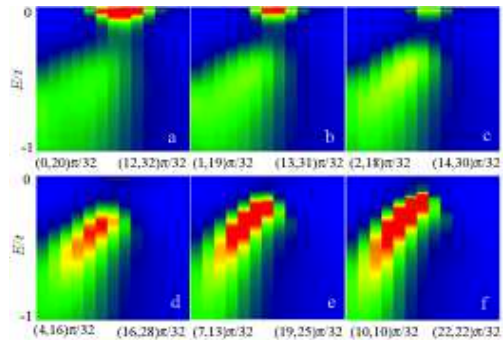, width=7. cm}}
\caption{(Color online) Occupied spectral density as function of momentum and energy at $x = 0.1$ and $T/t = 0.003$. The abscissa in (a)-(f) represent the momentum axis shown in Fig. \ref{fig14}.}\label{fig15}
\end{figure}

We here begin with the spectral density in the AF state and compare it with the experimental data. Exhibited in Fig. \ref{fig14} is the occupied spectral density around the chemical potential at $x = 0.1$ and $T/t = 0.003$. It shows that the Fermi surface is near the two points $(\pi,0)$ and $(0,\pi)$ where the spectral intensity is strong. This kind of Fermi surface has been observed in ARPES experiment by Armitage {\it et al.}\cite{Armitage} 
Later, Matsui {\it et al.}\cite{Matsui} obtained the momentum-energy-resolved result for the occupied spectral density by ARPES experiments. In order to compare our result with the corresponding experimental data, we show the occupied spectral density in Fig. \ref{fig15} as functions of momentum and energy. The momentum axes in Figs. \ref{fig15}(a)-(f) are along the lines a-f in Fig. \ref{fig14}. Figure \ref{fig15}(a) shows that the chemical potential cuts the upper band. The intensity in the region in the upper band close to the Fermi surface is very strong, illustrating the large weight of the quasiparticle there. On the other hand, the lower band is largely broadened, as shown by the wide area of weak intensity. From Figs. \ref{fig15}(a) to \ref{fig15}(f), the intensity distribution shifts upward gradually, and the lower band starts to show stronger intensity. In Fig. \ref{fig15}(c), the intensity is nearly vanishing at the chemical potential. This corresponds to the case of line c in Fig. \ref{fig14} where the line nearly misses to cross the
Fermi surface. In Fig. \ref{fig15}(d) there is no intensity at the chemical
potential since lind d as shown in Fig. \ref{fig14} passes approximately the hot spot. Comparing these figures, we find that the gap between the chemical potential and lower band increases gradually from Fig. \ref{fig15}(a) to Fig. \ref{fig15}(d) where the largest gap shows up. In Fig. \ref{fig15}(f), it is seen that the lower band is close to, but still away from the chemical potential. It should be pointed out that, at larger doping, the lower band will shift up and cross the chemical potential around the point $(\pi/2,\pi/2)$. Correspondingly, the spectral intensity near $(\pi/2,\pi/2)$ in Fig. \ref{fig14} will become strong enough and finally the several pieces of the Fermi surface will connect together. The present results are in fairly good agreement with the experimental observations.\cite{Armitage,Matsui} 

\begin{figure}
\centerline{\epsfig{file=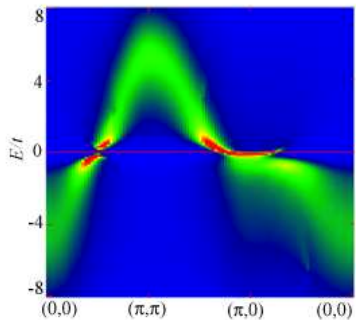, width=5.5 cm}}
\caption{(Color online) Spectral intensity as function of energy along the high-symmetry lines in the Brillouin zone at $x = 0.1$ and $T/t = 0.003$.}\label{fig16}
\end{figure}

\begin{figure}
\centerline{\epsfig{file=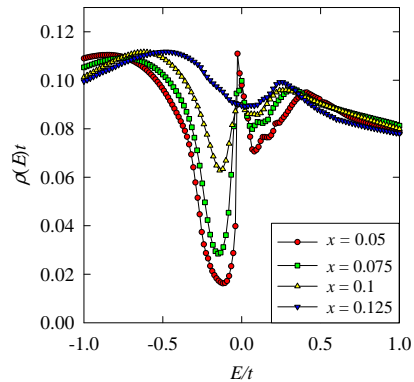, width=5.5 cm}}
\caption{(Color online) Density of states of AF phase at various doping concentrations at $T/t = 0.003$.}\label{fig17}
\end{figure}

In Fig. \ref{fig16}, we exhibit the spectral density along the high-symmetry lines in the Brillouin zone within a large energy scale to show the quasiparticle band structure at $T/t = 0.003$ and $x = 0.1$. In contrast to the mean field result, Fig. \ref{fig16} shows no clear dispersion relation between the energy and momentum, but considerable broadening of spectral density. Only close to the chemical potential the intensity shows sharp peak, implying that the quasiparticles have long life-time at the chemical potential. At $(\pi/2,\pi/2)$, the gap clearly shows up. The upper band is very flat at $(\pi,0)$, where it reaches its minimum. At low doping, the chemical potential is close to the bottom of the upper band. Even at zero doping, due to the many-body effect, the upper band is already occupied with quasiparticles; the chemical potential is at nearly the bottom of the upper band. From such a band structure, we can expect an unusual DOS. Shown in Fig. \ref{fig17} is the DOS of low temperature AF states at several doping concentrations. Clearly, a central peak appears in the DOS at zero energy at low doping, which originates from the fact that the chemical potential cuts the flat region near the point $(\pi,0)$ where the width of energy level is very narrow. With increasing the doping concentration the Fermi surface departures from the flat region, the peak in DOS at large doping is less pronounced. At $x = 0.125$, the peak diminishes as shown in Fig. \ref{fig17}, and it is expected to disappear at higher doping concentration. At zero doping, since the Fermi surface is very small, the zero-energy peak almost diminishes too. It should be noted that since the direct gap as appeared in the spectral density varies with the momentum, we obtain a structured gap in the DOS that is a superposition of the spectral density. We here call the distance between the humps below and above the chemical potential in the DOS as the gap. In Fig. 17, the gap appears between $E\sim -0.5t$ and $\sim 0.3t$. Especially, the bottom of the upper Hubbard band falls into the DOS gap consistent with the situation shown in Fig. \ref{fig16}. This is different from the MFT by which one obtains a DOS with a clear gap between the upper and lower bands, but not a peak at the chemical potential.

What is the physics of the structured gap with a central peak at zero energy in the DOS? In the system, there can be two kinds of elementary excitations: the single quasiparticles and the spin waves due to the transverse-spin fluctuations. The energy for the latter is within a range $0 \le \Omega \le \Omega_c$. Now, consider a quasiparticle with energy $E>0$ above the chemical potential. The quasiparticle can excite a spin wave by transferring energy $\Omega$ to the latter and transiting itself to a low energy $E-\Omega$ state. Such an excitation can happen in the energy range $0 \le \Omega \le {\rm min}(\Omega_c,E)$. Thus, the lifetime of the quasiparticle is finite, and the DOS at $E$ is suppressed. On the other hand, for the particle at very low energy, the energy range of the excited spin waves is very restricted; the integrated effect of the spin-wave excitations on the particle is very small. It means that the quasiparticles at very low energy have long lifetime. Therefore, the spin-wave excitations result in a dip in the DOS of the single particles above the chemical potential, but leave a peak at zero energy. The similar mechanism of the peak-dip-hump structure in the DOS of the HTS due to a dispersionless collective mode has been explained by Shen and Schrieffer.\cite{Shen-Schrieffer} It should be pointed out that besides the well-defined spin waves at low-energy regime, the spin fluctuations at higher energy can modify the DOS as well; the effects from both of them are included in the present calculation. The dip can be understood as pseudogap. The gap between the lower band and the chemical potential which is at about the bottom of the upper band corresponds to the mean field AF gap caused by the AF order. In the present case, it is rounded by the spin fluctuations. 

The present zero-energy peak in DOS may correspond to the in-gap states investigated by Kusunose and Rice.\cite{Kusunose} The origin of a in-gap state was explained as the admixture of the quasihole mean filed state in the lower band with the hole state accompanied by particle-hole excitations due to spin fluctuations in the upper band. By adopting the self-consistent Born approximation with the RPA spin-spin interaction and the MFT order parameter, they obtained the in-gap states at energies about $-t$ away from the chemical potential. In the present case, however, the order parameter, the spin fluctuations and the chemical potential adjust themselves all in a self-consistent way. As a result, the peak appears always at the chemical potential. 

\begin{figure}
\centerline{\epsfig{file=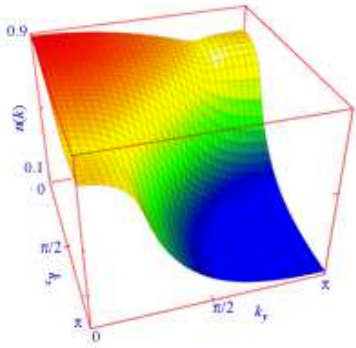, width=5.5 cm}}
\caption{(Color online) Distribution function $n(k)$ in the first quadrant of the Brillouin zone at $x = 0.1$ and $T/t = 0.003$.}\label{fig18}
\end{figure}

In Fig. \ref{fig18}, we depict the momentum-space distribution function $n(\vec 
k)$ (at $x = 0.1$) defined by
\begin{equation}
n(\vec k) = \frac{1}{\beta}\sum_nG(\vec k,i\omega_n)\exp(i\omega_n\eta)
\end{equation}
where $\eta \to 0^+$ is an infinitesimal small positive constant. Even at very low temperature $T/t = 0.003$, the distribution has a small but finite width without abrupt discontinuity at the Fermi surface, and thus is non-Fermi-liquid like. In addition, within the Fermi area, the states are not fully occupied in contrast to the case in a free electron system. Correspondingly, outside the Fermi area $n(k)$ is finite; the minimum of $n(k)$ is larger than 0.1. The distribution $n(k)$ shown in Fig. \ref{fig16} is a typical result within the doping range $0 \le x \le 0.17$. 

\begin{figure}
\centerline{\epsfig{file=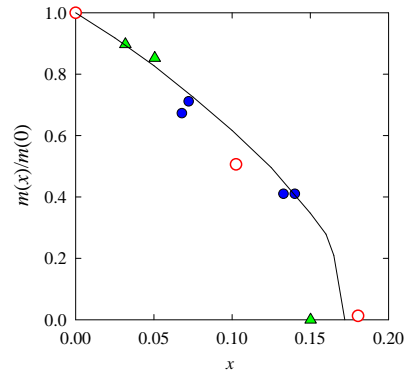, width=5.5 cm}}
\caption{(Color online) The calculated staggered magnetization $m$ (solid line) as a function of doping concentration $x$ at $T\approx 0 K$. The experimental data are also shown for comparison [solid and hollow circles,\cite{Mang} solid triangles\cite{Rosseinsky}].}\label{fig19}
\end{figure}

Figure \ref{fig19} exhibits the staggered magnetization $m$ as a function of $x$ at very low temperature $T \approx 0 K$, and the comparison with the experimental data.\cite{Mang,Rosseinsky} The experimental value for $m(0)$ at Cu$^{2+}$ is about 0.5 in NCCO and $\sim 0.4$ in PCCO,\cite{Thurston} which are close to our theoretical result 0.48. The overall behavior of $m$ versus $x$ obtained by the present calculation is in very good agreement with experiment. Though the dynamical-MFT calculation by S\'en\'echal {\it et al.} \cite{Senechal} yields a reasonable behavior for $m(x)/m(0)$, their result $m(0) \approx 0.7$ is too large. 

\section{Conclusion and remark}

In conclusion, we have presented the formulation of antiferromagnetism in electron-doped cuprate superconductors on the basis of the Hubbard model using the fluctuation-exchange approach. A self-consistent treatment of the spin fluctuations in combination with the impurity scattering due to the disordered dopant-atoms can reasonably account for a number of experimental results including the AF transition temperature $T_N$, the onset temperature of pseudogap $T^*$, the spectral density, and the staggered magnetization $m$. The single particle distribution function has also been calculated, and it differs significantly from that of the Fermi liquid. Also, we have found the existence of zero-energy peak in the density of states of the AF phase. The peak is sharp in the low doping regime, and disappears near the optimal doping. This result is not inconsistent with the ARPES experiments, but still needs to be carefully checked by the tunneling experiments.

Although the present theory provides reasonable explanations for a number of experiments, our calculated spin-wave velocity $c$, however, is smaller than the experimental data. With the hopping parameter $t \approx 270$ meV, the experimental results for $c$ are $0.74ta$ and $0.94ta$ respectively for PCCO and NCCO\cite{Bourges} ($a$: the lattice constant of CuO plane), while our theoretical prediction is only about $0.33ta$. The quantity $c$ is related to second order derivative of the transverse spin susceptibilities with respect to the momentum $d^2\chi^t_{\mu\nu}(Q+q)/dq^2 |_{q=0}$. In the present approach, we are probably able to reasonably treat the overall behavior of the two-particle propagator $\chi^t_{\mu\nu}(q)$, but not its second order derivative with respect to $q$. Further study is needed in this point.

\acknowledgments 

This work was supported by a grant from the Robert A. Welch Foundation under No. E-1146, the TCSUH, and the National Basic Research 973 Program of China under grant number 2005CB623602.

\vskip 5mm
\centerline {\bf APPENDIX}
\vskip 3mm

In this Appendix, we deal with the average of $V^t_{11}(q)$ over $-\pi<q_z<\pi$. As illustrated in Section II, we neglect the weak $q_z$-dependence in the numerator. Here, we consider only the case of $Q_z = 0$. The result for $Q_z = \pi$ can be derived analogously. 

For $\Omega_m = 0$, $\chi^t_{12}(q) = 0$, the expression for $V^t_{11}(q)$ reduces to the similar form as Eq. (\ref{vs}). Therefore, the average $\bar V^t_{11}(q)$ is given by 
\begin{equation}
\bar V^t_{11}(q) = 
\frac{U^2\chi^t_{11}(q)}{\sqrt{[1+U\chi^t_{11}(q)]^2-J^2_z\chi^{t2}_{11}(q)}}. \nonumber
\end{equation}

For $\Omega_m \ne 0$, the denominator of $V^t_{11}(q)$ that is the determinant $D = {\rm det}[1+\chi^t{q}V(q)]$ can be written as $D=a\cos^2q_z -b\cos q_z +c$ with  
\begin{eqnarray}
a &=& J^2_z[\chi^t_{11}(q)\chi^t_{22}(q)-\chi^{t2}_{12}(q)] \nonumber \\
b &=& J_z[\chi^t_{11}(q)+\chi^t_{22}(q)]+2aU/J_z \nonumber \\
c &=& 1+U[\chi^t_{11}(q)+\chi^t_{22}(q)]+a(U/J_z)^2. \nonumber 
\end{eqnarray}
Notice that $\chi^{t}_{12}(q)$ is a pure imaginary quantity. $D$ never vanishes at $\Omega_m \ne 0$. By denoting the average by 
\begin{equation}
\bar V^t_{11}(q)=U^2(1+aU/J_z^2)g(q), \nonumber 
\end{equation}
the function $g(q)$ is given by

(A)	$d=b^2-4ac > 0$,
\begin{equation}
g(q) = \frac{1}{\sqrt{d}}[{\rm  
sgn}(x_-)\frac{\theta(|x_-|-1)}{\sqrt{x^2_--1}}
-	{\rm sgn}(x_+)\frac{\theta(|x_+|-1)}{\sqrt{x^2_+-1}}]\nonumber
\end{equation}
where $x_{\pm}=(b\pm\sqrt{d})/2a$.

(B)	$d= 0$,
\begin{equation}
g(q) = \frac{|x|}{a(x^2-1)^{3/2}}\nonumber
\end{equation}
where $x = b/2a$.

(C) $d < 0$,
\begin{equation}
g(q) = -\frac{2}{\sqrt{|d|}}{\rm Im}(z^2-1)^{-1/2}\nonumber
\end{equation}
where $z=(|b|+i\sqrt{|d|})/2a$.

\end{document}